# EELGRASS BEDS AND OYSTER FARMING IN A LAGOON BEFORE AND AFTER THE GREAT EAST JAPAN EARTHQUAKE OF 2011: POTENTIAL FOR APPLYING DEEP LEARNING AT A COASTAL AREA

*Takehisa Yamakita*

Marine Biodiversity and Environmental Assessment Research Center (BioEnv),
Japan Agency for Marine-Earth Science and Technology (JAMSTEC)

## ABSTRACT

There is a small number of case studies of automatic land cover classification on the coastal area. Here, I test extraction of seagrass beds, sandy area, oyster farming rafts at Mangoku-ura Lagoon, Miyagi, Japan by comparing manual tracing, simple image segmentation, and image transformation using deep learning. The result was used to extract the changes before and after the earthquake and tsunami. The output resolution was best in the image transformation method, which showed more than 69% accuracy for vegetation classification by an assessment using random points on independent test data. The distribution of oyster farming rafts was detected by the segmentation model. Assessment of the change before and after the earthquake by the manual tracing and image transformation result revealed increase of sand area and decrease of the vegetation. By the segmentation model only the decrease of the oyster farming was detected. These results demonstrate the potential to extract the spatial pattern of these elements after an earthquake and tsunami.

## 1. INTRODUCTION

In recent years, aerial images have become available at finer resolution and lower prices, leading to new applications of remote sensing to perform object detection and recognition that are beyond pixel color-based classification and extraction. Automated analysis using deep learning is expected to allow faster, effortless, and more accurate analysis of images [1]. This is especially useful under urgent circumstances, such as monitoring following a disaster such as an earthquake and tsunami [2].

Compared to the increasing number of applications of automatic recognition technology to terrestrial remote sensing, applications to coastal and marine areas are not popular. These have been especially limited in the recognition of ecological and biological objects [3]. However, among such applications, the recent application of automatic image transformation techniques to seagrass has demonstrated the feasibility of extracting seagrass bed accurately and within a few minutes from even old grayscale images [4]. This model, which uses deep learning, can detect the texture, shapes, and contrast of underwater seagrass vegetation. Thus, this method may also be applicable for other land uses or objects distinguishable by shape and texture.

For rapid response to disaster and management of a coastal area, it is necessary to extract more types of elements from images. In particular, consideration of the distributions of oyster farming rafts, and debris are essential for studying the tsunami damage that occurs periodically in Japan. However, existing image datasets, such as ImageNet, cover a very limited number and types of data on marine or remote sensing elements [5], [6]. Not only tsunami-related coastal elements but also elements necessary for performing aerial photograph analysis are lacking in public databases. Therefore, it is also necessary to consider creating supervised data for the extraction of these elements.

In the aforementioned paper on seagrass extraction [4], the accuracy of the new method was compared with existing pixel color-based classification methods in seagrass remote sensing research [7], [8]. However, the paper did not compare the accuracy between the different models of deep learning. As representative deep learning models for an image, a model to identify the object, a model to fill the location of the object, and a model to convert an image to another image having the same characteristics have been proposed. However, comparisons of these different types of models have been conducted mostly for popular benchmark datasets, which are not marine remote sensing data. Therefore, it is necessary to compare the effects of the characteristics of the different types of models on the classification results of realistic data of coastal and marine remote sensing.

Here, I propose an appropriate automatic detection method of seagrass beds, oyster farming rafts (also called oyster shelves), and marine debris using aerial images. Additionally, I apply the model to a comparison of the distributions of those land cover types before and after the Great East Japan

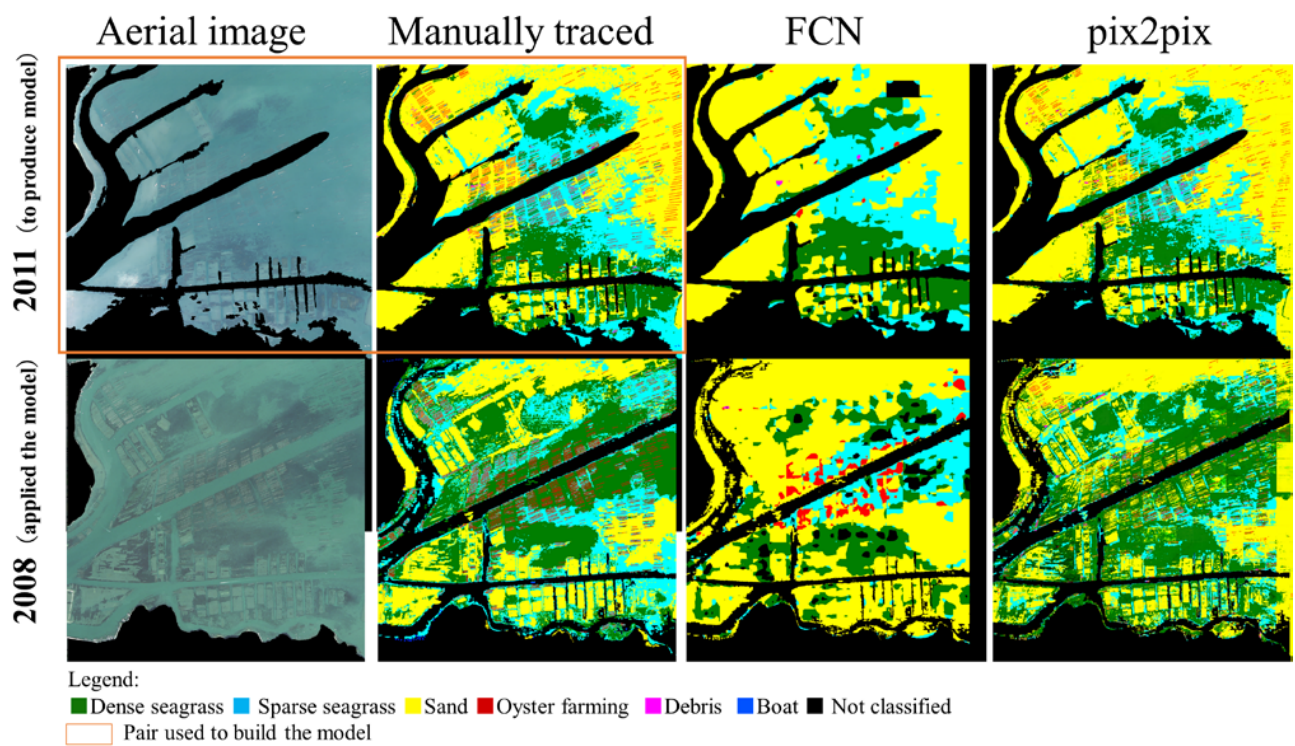

Fig. 1. Original images and classified images of the Mangoku-ura Lagoon, Miyagi, Japan. Images taken in 2008 and 2011 were used to represent before and after the earthquake. The pair of images in the orange rectangle was used to build the models using fully convolutional network (FCN) and pix2pix which implement Deep Convolutional Generative Adversarial Network (DCGAN).

Earthquake and subsequent tsunami in 2011. I especially tested the image segmentation, and image-to-image conversion (translation).

## 2. METHODS

As the study site, I selected Mangoku-ura Lagoon, near the Oshika Peninsula, within the region damaged by the Great East Japan Earthquake. The lagoon, which is one of the largest enclosed bodies of water in this region, includes a wide distribution of seagrass bed and oyster farming areas [9]–[11]. Because of the earthquake, this area was affected by a 0.7 m ground subsidence and a 7.7 m tsunami [12], [13]. The west side of the lagoon, which is near the mouth of the lagoon and may have suffered greater effects from the tsunami, was selected as the area of focus.

Aerial images acquired on Oct. 13, 2008, and Jun. 7, 2011, were available with clear transparency of the water (Fig. 1). All images were resampled to the same resolution as that of the 2008 imagery (ca. 40 cm per pixel) and were adjusted to the same color level of 2011 imagery before the analysis. The image acquired in June 2011 was used to create supervised data for the modeling. As the training data which common to all methods, manual extraction of objects and land cover was conducted by persons including those who participated in the field survey [4].

For the automatic extraction model, I utilized two representative techniques of deep learning. According to the order of model complexity, the following models were used. A model to identify an object using the convolutional neural network (CNN) was the most basic model [14], [15]. By modifying the CNN, it is also possible to fill the location of objects from a large-size image. The modification is basically conducted by exchanging the fully connected layer, which returns the probability of the identification class to the convolutional layer, which shows a two-dimensional map of the probability. This image segmentation method was named "fully convolutional networks" (FCNs).

Recently, image conversion using networks that produces an image using the features extracted by normal CNN have become popular. This model is intended to allow segmentation together with the functionality to reproduce images. The model trains the network that generates the image by competing with another CNN that compares the similarity of the output image to the supervised data. Because of that competition, this image-to-image conversion

TABLE I. THE ACCURACY OF THE RESULT USING RANDOM POINTS AGAINST MANUAL CLASSIFICATION.

| Year | Model | Accuracy | | | | Producer's Accuracy (recall, sensitivity) | | | | | | | User's accuracy (precision) | | | | | | | Accuracy (rand accuracy) | | | | | | |
|---|---|---|---|---|---|---|---|---|---|---|---|---|---|---|---|---|---|---|---|---|---|---|---|---|---|---|
| | | Total vegetation | | Two vegetation classes | | | Vegetation | | | Oyster | | Not | | Vegetation | | | Oyster | | Not | | Vegetation | | | Oyster | | Not |
| | | Overall | Kappa | Overall | Kappa | Sand | Total | Dense | Sparse | rafts | Debris | Classified | Sand | Total | Dense | Sparse | rafts | Debris | Classified | Sand | Total | Dense | Sparse | rafts | Debris | Classified |
| 2011 | FCN | 88% | 0.84 | 85% | 0.79 | 91% | 81% | 79% | 63% | 2% | 10% | 99% | 85% | 88% | 79% | 75% | 31% | 36% | 93% | 91% | 90% | 94% | 90% | 98% | 100% | 98% |
| | pix2pix | 91% | 0.87 | 88% | 0.83 | 92% | 87% | 83% | 74% | 17% | 33% | 98% | 88% | 89% | 81% | 80% | 26% | 42% | 98% | 93% | 92% | 95% | 92% | 98% | 100% | 99% |
| 2008 | FCN | 58% | 0.45 | 49% | 0.33 | 87% | 39% | 28% | 13% | 14% | 0% | 78% | 45% | 81% | 59% | 27% | 20% | 0% | 67% | 69% | 64% | 72% | 75% | 95% | 0% | 88% |
| | pix2pix | 67% | 0.56 | 55% | 0.40 | 63% | 69% | 62% | 20% | 3% | 0% | 80% | 55% | 73% | 52% | 34% | 13% | 0% | 75% | 77% | 71% | 71% | 76% | 95% | 0% | 91% |

*Total of vegetation means the result after addition of dense seagrass and sparse seagrass categories
*Rand accuracy = (TP+TN)/(TP+TN+FP+FN)
TP = True positive; FP = False positive; TN = True negative; FN = False negative

TABLE II. LANDCOVER CHANGE BY DIFFERENT METHODS

| | | | Vegetation types | | | Oyster | | Not |
|---|---|---|---|---|---|---|---|---|
| | | Sand | Total | Dense | Sparse | rafts | Debris | classified |
| 2011 | Manually | 101.9 | 87.8 | 39.8 | 48.0 | 4.9 | 0.5 | 75.1 |
| | FCN | 109.4 | 80.0 | 39.9 | 40.1 | 0.4 | 0.2 | 80.3 |
| | pix2pix | 106.4 | 85.2 | 41.3 | 44.0 | 3.3 | 0.5 | 74.9 |
| 2008 | Manually | 70.6 | 137.5 | 82.1 | 55.4 | 10.8 | 0.0 | 51.4 |
| | FCN | 135.9 | 66.9 | 39.0 | 27.9 | 7.4 | 0.0 | 60.0 |
| | pix2pix | 79.5 | 132.7 | 99.2 | 33.5 | 2.7 | 0.7 | 54.7 |

(ha)

method was named "deep convolutional generative adversarial network" (DCGAN).

To use these models, standard network structures are applied in each type of model. I used the FCN version of AlexNet as the FCN [14],[16], and the pix2pix [17] to implement DCGAN, respectively.

For the segmentation and image conversion model, aerial images and the classified image for supervised data were sliced into 256 pixel grids. A pair of sliced supervised images and aerial images of the same year was produced to feed the model. To apply the model, the remaining aerial image was used as test data. After classification by the model, the sliced images were merged into the same location of the original image.

To evaluate the accuracy of the results, I compared the classification results by the models and the manually classified result pixel-by-pixel using more than 100,000 random points under the ArcGIS 10.2. At that time, areas that could not be classified because of the boundary of the image were masked in advance. The pixels classified with low probability by FCN and the color noise generated by pix2pix were also integrated into the category of the nearest color class in advance.

Trends of change during the years acquired from the aerial images were evaluated by the relative change in the pixel number of each category. The results of areal changes of seagrass and sand were qualitatively compared with past research that focused on the entirety of this lagoon [10], [11].

## 3. RESULTS

Fig. 1 shows the original images and the result of the classification. The accuracy of the result is shown in Table I. In the rows labeled "2011", the result indicates the agreement with the model output against the supervised data used to produce the model. There was 80 % accuracy overall and more than 60 % agreement in the categories of sand, dense vegetation, sparse vegetation and the total of both vegetation in both models. The agreement against the independent manually classified data is shown in the rows named "2008". There was 67 % accuracy overall (when considering total vegetation) and more than 60 % agreement by the producer's accuracy in the categories of sand, dense and total vegetation in the model using pix2pix. However, the overall accuracy was 58 % and the agreement was lower than 40 % by the producer's accuracy in the categories of vegetation and oyster farming rafts in the FCN model. In addition, many locations were not classified, as shown in Fig. 1, by the FCN model.

For the classification of oyster farming rafts, the shapes of the shelves or rafts were clearly categorized even for the test data in the model using pix2pix, as observed at the center of the images in Fig. 1 (2008 pix2pix). However, the classified pixels were not always oyster farming. Some were classified as sandy area and some were classified as debris. This may have been caused by exposure of the structures on the water before the disaster accompanied by ground subsidence. In contrast, the areas of oyster farming were correctly classified in the case of the test data 2008 imagery by the FCN model. It may because of the advantage of the larger window size of FCN to recognize objects. However, the resolution of the model using FCN was too low to classify shapes of oyster farming rafts.

The result of the change from pre-earthquake status (Table II) shows decreases of dense seagrass beds and oyster farming rafts and increases of sand and marine debris. Similar trends were observed using pix2pix for sand and dense vegetation. Using the FCN, similar aerial change of the oyster farming rafts was detected. This difference may have been caused by the feature of the model, contamination of fragments of noise at the edge of the analysis grids, and the change of water depth.

## 4. CONCLUSION

In this study, we proposed methods for extracting seagrass beds and oyster farming rafts from aerial photographs using automatic image analysis techniques applying deep learning. The comparison of the two techniques revealed that the pix2pix technique produces higher accuracy for the vegetation extraction. The resolution of the result image is also high with this model and extracted the shape of the oyster farming rafts. Categorization of areas of oyster farming rafts was better to use the FCN in this data. This categorization may also potentially possible after post-processing of categories from the result of pix2pix. From the results of the comparison before and after the earthquake and tsunami, the number of boats, oyster farming rafts, and seagrass beds decreased. To show this pattern, use of the appropriate model, consideration of water depth and automatic reduction of noise in the models is needed in the future. As shown here, these results demonstrate the potential for quick detection of changes in the coastal area following an earthquake by this technique.

## 5. ACKNOWLEDGMENT

I thank Fisherman's Association of Miyagi Prefecture, Miyagi Prefecture Fishery Technology Institute, Masakazu Hori, Daisuke Muraoka, Goro Yoshida, Hitoshi Tamaki, and other members of the Fisheries Research and Education Agency (FRA) for the opportunity of field observation. I also would like to thank Tomonori Mita (SKYMAP Co., Ltd.) and Fumiaki Sodeyama (JAMSTEC) for technical assistance. A portion of this research was supported financially by the Tohoku Ecosystem-Associated Marine Sciences (TEAMS) research program funded by the Ministry of Education, Culture, Sports, Science and Technology (MEXT) and the Environment Research and Technology Development Fund (ERTDF, S-15 PANCES Project) of the Ministry of the Environment, Japan.